\documentclass[%
 aip,
 jcp,
 amsmath,amssymb,
preprint,
]{revtex4-1}

\usepackage{graphicx}% Include figure files
\usepackage{dcolumn}% Align table columns on decimal point
\usepackage{bm}% bold math
\usepackage[utf8]{inputenc}
\usepackage[T1]{fontenc}
\usepackage{mathptmx}
\usepackage{etoolbox}

\makeatletter
\def\@email#1#2{%
 \endgroup
 \patchcmd{\titleblock@produce}
  {\frontmatter@RRAPformat}
  {\frontmatter@RRAPformat{\produce@RRAP{*#1\href{mailto:#2}{#2}}}\frontmatter@RRAPformat}
  {}{}
}%
\makeatother
\begin{document}

\preprint{AIP/123-QED}

\title{Liquid--vapor critical behavior of the TIP4P/2005 water model: effects of NaCl solutes and hydrophobic confinement}
%\title{Influence of ionic condiditions and confinements on the liquid-gas critical point of the TIP4P/2005 water model}
\author{Mayank Sharma}
\affiliation{Johannes Gutenberg-Universität, Institut für Physik, Mainz, Germany}
\author{Peter Virnau}
\email{virnau@uni-mainz.de}
\affiliation{Johannes Gutenberg-Universität, Institut für Physik, Mainz, Germany}
\date{\today}% It is always \today, today,

\begin{abstract}
The liquid--vapor critical behavior of water is strongly influenced by both ionic solutes and confinement. Molecular dynamics simulations of aqueous NaCl solutions using the TIP4P/2005 water model and the Madrid-2019 ion parameters reveal a systematic increase in the liquid--vapor critical temperature and pressure with salt concentration, consistent with experimental trends. In contrast, confinement between parallel hydrophobic plates leads to a depression of the critical point. \
The critical temperature was determined from the Binder cumulant crossing in the NVT ensemble, using a recently developed method originally applied to an active Brownian particle system. The reliability of this approach was verified through complementary NPT simulations. We further demonstrate the pronounced sensitivity of the estimated critical point to the van der Waals cutoff distance, underscoring the importance of properly accounting for long-range interactions. Our framework, which employs and tests an adjusted subbox method, can be readily extended to other critical phenomena, including the liquid--liquid critical point of water.

\end{abstract}

\maketitle
\pagebreak

%\section{\label{sec:level1}Introduction\protect\\ The line
%break was forced \lowercase{via} \textbackslash\textbackslash}
\section{\label{sec:level1}Introduction}

The liquid--vapor critical point is a defining thermodynamic property that has played a central role in the development of modern statistical physics since the 19th century with wide-ranging importance, from climate science to industrial processes \cite{andrews1869xviii, stillinger1980water, debenedetti2020metastable, weingartner2005supercritical}. In bulk water, the critical point occurs at $T_c \approx 647.1$\,K and $P_c \approx 22.06$\,MPa \cite{wagner2002iapws}. Dissolved salts substantially alter the phase behavior; for example, seawater ($\approx$3.2 wt\% NaCl) boils at higher temperatures and its critical point is shifted relative to pure water \cite{bischoff1984critical, marshall1974liquid, bischoff1988liquid, pitzer1984critical}. Understanding how electrolytes and nanoscale confinement modify the critical behavior of water is relevant to geological systems, supercritical reactors, and other materials applications \cite{bischoff1984critical,weingartner2005supercritical,marrone2013supercritical}. Confinement effects are particularly significant: Water confined in highly hydrophobic nanopores exhibits a critical temperature up to several hundred Kelvin lower than in bulk water\cite{merchiori2024mild, beckstein2003liquid}. The putative liquid–liquid critical point for water has motivated extensive research into how water’s critical behavior responds to external perturbations\cite{poole1992phase, xu2005relation, franzese2002liquid, franzese2007widom, neophytou2022topological, harrington1997liquid, kesselring2012nanoscale, palmer2018advances, sciortino1997line, poole2005density, gallo2010dynamic, corradini2010route, stokely2010effect, gallo2017supercooled, franzese2001generic, kumar2008predictions}. Subsequent studies have shown that this behavior is highly sensitive to solutes and confinement, highlighting the complex interplay between water’s structure and its environment \cite{bianco2014critical, gallo2016water, gallo2021advances, kumar2005thermodynamics, corradini2011liquid, debenedetti2020second, yagasaki2019liquid, knight2019water, longinotti2011anomalies, han2009hydrogen, romero2009effect, srivastava2011phase}.

%Similar sensitivity of water’s critical behavior to solutes and confinement has also been discussed in the context of the putative liquid--liquid critical point\cite{bianco2014critical, gallo2016water, kumar2005thermodynamics, corradini2011liquid, debenedetti2020second, yagasaki2019liquid, knight2019water, longinotti2011anomalies, han2009hydrogen, romero2009effect, srivastava2011phase, sciortino1996supercooled, gallo2010dynamic}.

Molecular dynamics (MD) simulation has long been used to connect microscopic interactions to macroscopic phase behavior in water\cite{binder2004molecular, rahman1971molecular, stillinger1974improved, hansson2002molecular, hoover1986molecular, baran2022ice}. Historically, rigid non-polarizable models such as TIP3P\cite{jorgensen1983comparison}, SPC/E\cite{berendsen1987missing}, and TIP4P\cite{jorgensen1983comparison} were developed to reproduce key properties of liquid water, including density, diffusion, and enthalpy of vaporization, albeit with limitations in accurately reproducing phase equilibria \cite{vega2011simulating, vega2009ice}. Subsequent improvements, including TIP4P/Ew\cite{horn2004development} and TIP4P/2005\cite{abascal2005general}, were designed to better match the temperature of maximum density (TMD), melting points, and vapor--liquid coexistence curves while maintaining computational efficiency \cite{vega2011simulating}. Among these, TIP4P/2005 is widely used for critical-point studies because it reproduces vapor--liquid equilibria more accurately than three-site and some other four- or five-site potentials \cite{abascal2005general, vega2006vapor, hu2015liquid}. Accurately locating critical parameters in such models relies on methodological advances originally developed in Monte Carlo studies~\cite{landau2021guide,newman1999monte}, including finite-size scaling
~\cite{wilding1995critical,wilding1992density,bruce1992scaling,wilding1997simulation,binder1984finite,binder1987finite}, Binder-cumulants ~\cite{rovere1990gas,binder1981finite,binder1992monte}, and histogram reweighting schemes~\cite{ferrenberg1988new,ferrenberg1989optimized}.
These approaches substantially reduce finite-size biases that can obscure the critical point~\cite{siebert2018critical,virnau2002phase}.

Developing reliable models for aqueous electrolyte solutions requires careful consideration of the water--water, ion--ion, and ion--water interactions. Early studies relied on implicit-solvent models, and non-polarizable water models such as SPC/E remain widely used, although the choice of water model may affect how the remaining interactions are optimized\cite{benavides2017potential, kann2014scaled}. Given the limitations of SPC/E in reproducing key anomalies such as the temperature of maximum density\cite{vega2011simulating}, later efforts focused on constructing ion models compatible with TIP4P/2005\cite{kann2014scaled, benavides2017potential}. Among these developments, the Madrid-2019 force field provides a consistent and widely adopted framework for modeling monovalent and divalent ions in combination with TIP4P/2005\cite{zeron2019force}. It has also been shown to outperform earlier non-polarizable models across a wide range of properties, including solution densities over a wide concentration range, stability near experimental solubility limits without spurious crystallization, and realistic behavior in binary and ternary mixtures\cite{habibi2022new, blazquez2023scaled}. Extensions of this framework to divalent cations such as Sr$^{2+}$, Ba$^{2+}$ , and other ions have further broadened its applicability to more complex ionic systems\cite{blazquez2022madrid, blazquez2024madrid, trejos2023further, lamas2022freezing}. Moreover, recent experimental determinations of the temperature of maximum density (TMD) for a variety of salts show that the Madrid-2019 force field accurately reproduces the corresponding TMD shifts in aqueous electrolyte solutions\cite{sedano2022maximum, blazquez2023scaled}. This provides additional support for its ability to capture subtle thermodynamic anomalies in salt--water mixtures.

Despite extensive research on water’s interfacial and supercritical properties, systematic molecular dynamics studies directly quantifying how ionic solutes or hydrophobic confinement modify its liquid--vapor critical behavior remain scarce. In particular, only limited simulation work exists for confinement effects\cite{merchiori2024mild, brovchenko2001phase, brovchenko2004water, brovchenko2004water2, striolo2005water}, and some investigations have considered ionic effects on the liquid--liquid critical point\cite{la2025molecular, corradini2011liquid, yagasaki2019liquid, longinotti2011anomalies}. Few simulation studies have explored NaCl--water mixtures at subcritical and supercritical conditions and compared modern ion models, including Madrid-2019, with other force fields\cite{maerzke2022monte, patel2021nacl}. These studies characterize portions of the vapor--liquid coexistence region but do not extract the liquid--vapor critical point from simulation. Ref.~\onlinecite{patel2021nacl} has also reported qualitative indications consistent with salt-induced increases in the apparent vaporization temperature, often discussed in the context of the Driesner equations of state\cite{driesner2007system, driesner2007system2}; however, that analysis relies on density discontinuities rather than determining the critical point from simulation, leaving the concentration dependence of the critical point unquantified. Moreover, the sensitivity of simulated critical parameters to the treatment of dispersion (finite van der Waals cutoffs versus reciprocal-space treatments such as LJ-PME) requires careful validation to separate numerical artifacts from physical trends \cite{smit1992phase, hu2015liquid, klauda2007long, qiu2019systematic, miguez2013influence}. 

Here, we address these gaps using the TIP4P/2005 water model. We perform NVT and NPT simulations across multiple system sizes, employing a novel variant of the subsystem Binder-cumulant method, histogram reweighting, and finite-size scaling to estimate the critical temperature in the thermodynamic limit\cite{debenedetti2020second, ferrenberg1989optimized, tsypin2000probability}. We also systematically assess the effect of the Lennard-Jones cutoff compared with LJ-PME treatments\cite{hu2015liquid}. Our results show that the treatment of long-range dispersion has a pronounced influence on the apparent critical parameters, with LJ-PME eliminating the residual cutoff dependence that persists even when analytical long-range Lennard-Jones corrections are applied. For aqueous NaCl solutions and water confined between parallel hydrophobic plates, we use NVT simulations together with the Binder-cumulant method to quantify the influence of ionic solutes and confinement on the liquid--vapor critical behavior. We find that dissolved NaCl (Madrid-2019 parameters) elevates the critical temperature, whereas hydrophobic confinement depresses it, in qualitative agreement with prior studies\cite{brovchenko2004water, merchiori2024mild, rouault1995phase, vink2006phase, binder2008confinement, liu2010finite}. These contrasting trends highlight distinct microscopic mechanisms by which solutes and confinement modify the liquid--vapor critical region.

\section{Methods}

Accurate direct determination of critical points with Monte Carlo simulations of the Ising model were pioneered by Binder in the 1980s by combining subbox block distribution methods with cumulant crossings~\cite{rovere1990gas,binder1981finite,binder1992monte}. 
%
%While conceptually powerful, this approach has several limitations: requiring large system sizes, lacking a direct relation between the full-system correlation length and subbox size, and, in the NVT ensemble, dominant interfacial effects leading to no cumulant crossings in 2D Ising model\cite{rovere1993simulation,siebert2018critical}. 
%
Instead of determining cumulant crossings, order-parameter distributions can also be mapped onto the universal Ising master curve after proper normalization\cite{privman1984universal,ferrenberg1988new} for systems belonging to the Ising universality class. Nowadays, direct determination of critical points is most reliably achieved using grand canonical Monte Carlo simulations combined with field-mixing and histogram reweighting, following an approach developed by Bruce and Wilding in the 1990s\cite{bruce1992scaling,wilding1992density,wilding1995critical,wilding1997simulation}. This method is particularly advantageous as it directly relates the size of the simulation box with the correlation length of the system.
In principle, simulations in the isothermal–isobaric (NPT) ensemble also yield density-dependent probability distributions circumventing the problem of inserting complex molecules into dense phases~\cite{conrad1998comparison,debenedetti2020second}. In this work, we test this approach and quantify first-order finite-size effects arising from mapping the density distribution onto the Ising master curve. We also consider an improved subbox method~\cite{siebert2018critical}, which avoids interfacial biases and the requirement of large system sizes in the NVT ensemble \cite{rovere1993simulation,siebert2018critical} and is (just as the previous method) applicable to molecular dynamics simulations using standard simulation packages.

To accurately determine critical points, the methodology employed in this study closely follows the slab geometry block distribution scheme originally introduced in Ref.~\onlinecite{siebert2018critical}, which was applied to active systems in two dimensions. This approach modifies the original block-density distribution methods\cite{binder1981finite, rovere1988block, rovere1990gas, watanabe2012phase}, in which a square system is replaced by an elongated box with dimensions \(L_x = 6l\) and \(L_y = 2l\), promoting a slab-like configuration that stabilizes planar interfaces between the liquid and gas phases. The densities are then sampled by placing two subboxes (each with side length \(l\)) at the system’s center of mass and the other two at a position shifted by \(3l\) along the \(x\)-direction thus sampling only one third of the simulation box volume. This arrangement ensures that the subboxes probe the bulk liquid and gas regions while avoiding contributions from the interfacial region. Thereby, we enable accurate sampling of density fluctuations similar to those obtained for grand canonical Monte Carlo simulations \cite{wilding1992density, wilding1995critical,siebert2018critical,virnau2002phase} while preserving the accessibility of an MD-based approach. Furthermore by adjusting $l$, we can capture the finite-size dependence of the correlation length directly. 

\begin{figure}[t]
  \centering
  \begin{minipage}[c]{0.45\linewidth}
    \centering
    \includegraphics[width=\linewidth]{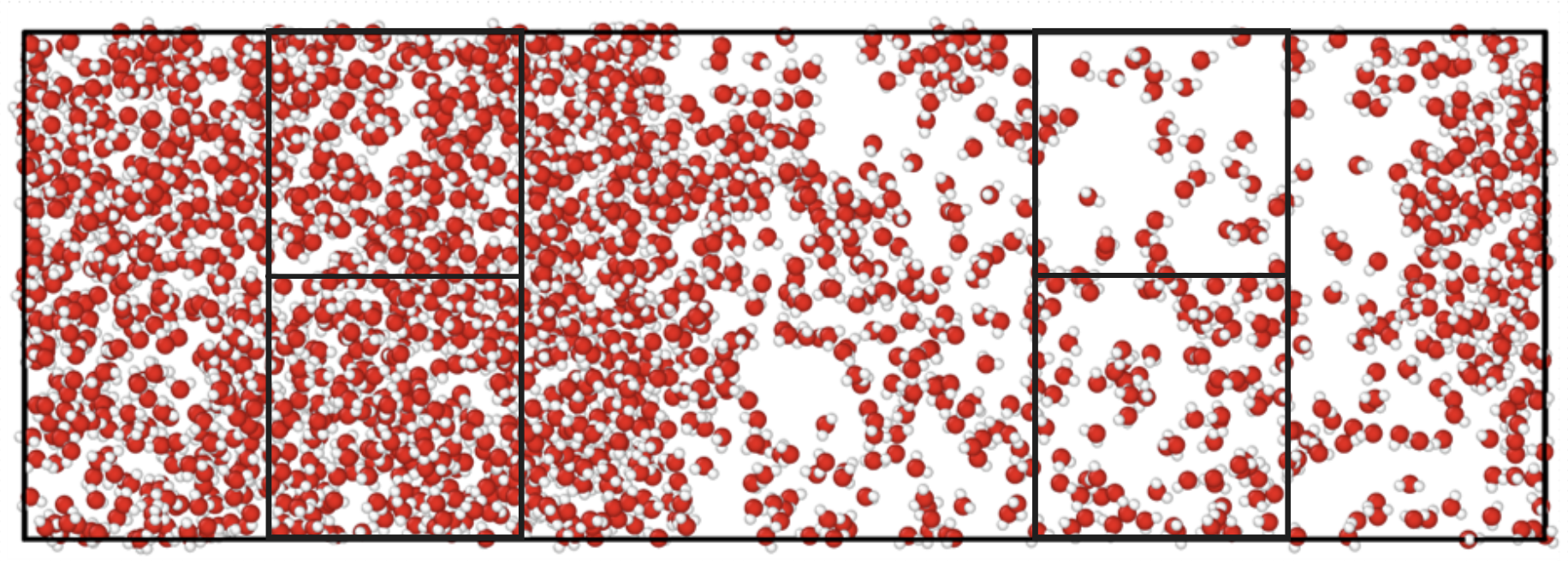}
    \par\smallskip
    (a)
  \end{minipage}
  \hfill
  \begin{minipage}[c]{0.45\linewidth}
    \centering
    \includegraphics[width=\linewidth]{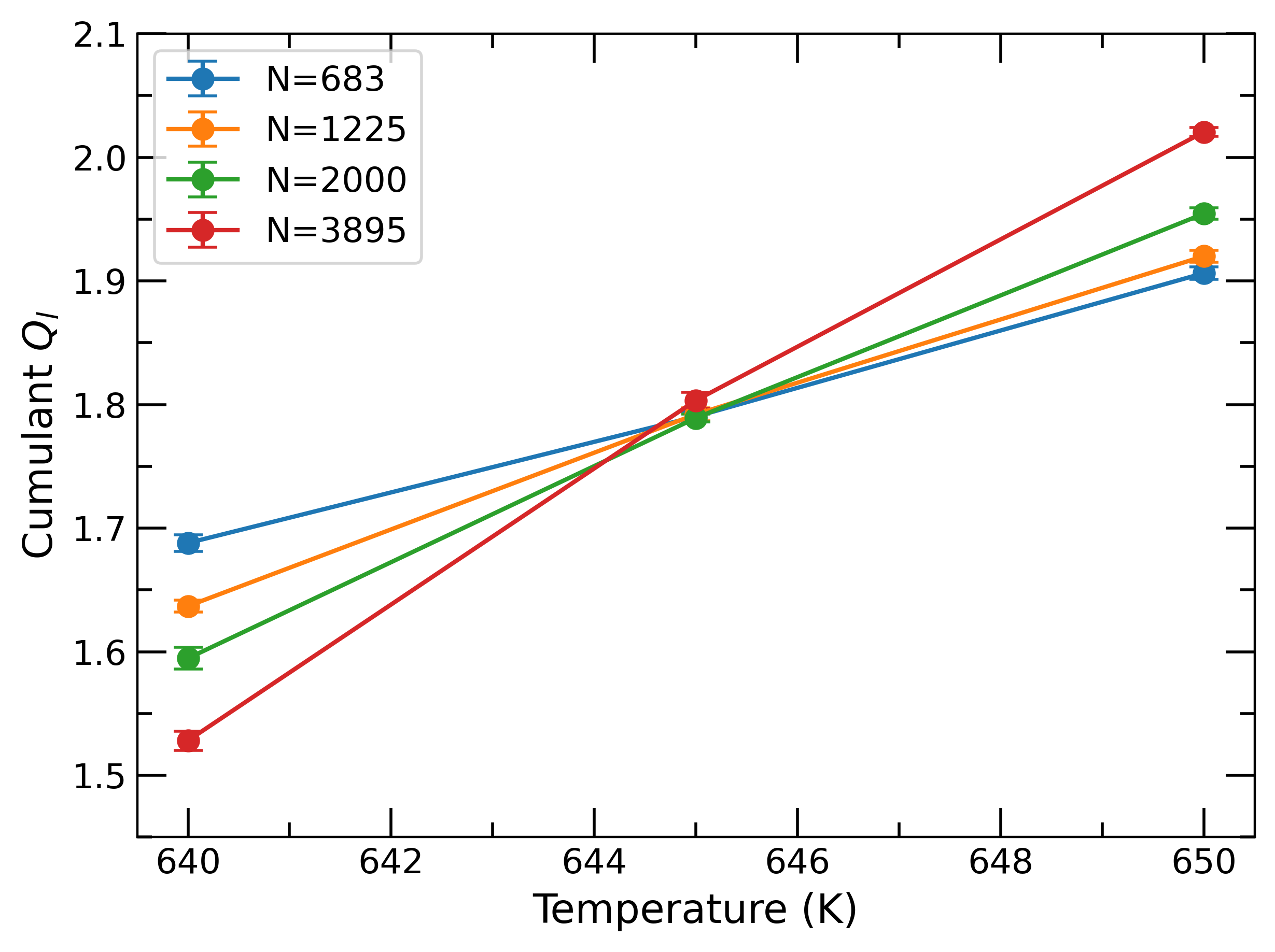}
    \par\smallskip

    (b)
  \end{minipage}
  \caption{
        (a) Simulation setup used for the Binder-cumulant analysis. Cubic subboxes of side length $l$ are placed in regions of maximum (liquid) and minimum (gas) number density inside a simulation box of dimensions $6l \times 2l \times 2l$. Only four out of eight subboxes are visible in the 2D $xy$ projection shown here.
        (b) Binder-cumulant crossings for four system sizes (with $l = 1.4,~1.7,~2.0,~2.5$~nm). The intersection of the curves between $N=2000$ and $N=3895$ indicates a critical temperature of approximately 644.3 K.
    }
  \label{fig:1} 
\end{figure}

In this study, we adapt this scheme to water system by using a three-dimensional elongated box (\(L_x = 6l\), \(L_y = 2l\) and \(L_z = 2l\)). To reduce drift effects in the wrapped trajectories near the critical point, we divide the simulation box into six slabs along the \(x\)-axis, each with a thickness of \(l\) (Figure~\ref{fig:1}a). Subboxes are then sampled only within slabs that exhibit the highest and lowest local densities, ensuring subbox placement within well-defined bulk regions away from the interface. A total of eight cubic sub-regions, each with a side length of \(l\), are analyzed in each configuration.

The order parameter is defined as: 
\begin{equation}
    M_i = \rho_i - \langle \rho \rangle
\end{equation}
where \(\rho_i\) is the density in a particular subbox and \(\langle \rho \rangle\) being the average density in the subboxes. We define the Binder cumulant as:
\begin{equation}
    Q_l = \frac{ \langle M_l^4 \rangle} {\langle M_l^2 \rangle^2}
\end{equation}
where the quantity \(Q_l\) depends on the system size \(l\). 
To account for finite-size effects and enable finite-size scaling analysis, simulations were performed for a range of subbox sizes \(l\). By evaluating the binding cumulant \(Q_l\) across different values of \(l\), the critical point can be identified as the temperature at which the curves corresponding to various \(l\) intersect (Figure~\ref{fig:1}b). This crossing behavior is a hallmark of criticality and allows a reliable determination of the critical parameters in finite systems. In practice, finite-size correction and higher-order scaling terms prevent all curves from crossing at a single point, resulting in small inconsistencies between system sizes~\cite{ferrenberg2018pushing}. Smaller system sizes exhibit the strongest finite-size effects, and therefore their crossing temperatures tend to show the largest deviations from the true critical temperature $T_c$. For simplicity, the crossing between the two largest system sizes is used here, although a residual finite-size shift on the order of 1 K may persist.

For the confined water simulations, we modeled the walls using a purely repulsive potential of the form
\begin{equation}
    U(z) = \left(\frac{1}{z+0.9}\right)^{200},
\end{equation}
which diverges steeply at short distances. The potential was truncated at $r_c = 0.2$~nm and shifted such that $U(r_c) = 0$, effectively acting as a hydrophobic wall. This choice minimizes depletion effects that can arise when Lennard-Jones type wall potentials are used\cite{kumar2007effect}. In practice, the walls act as ideal hydrophobic boundaries, simply excluding water molecules from penetrating beyond the surface. The simulation box geometry was $L_x=6l, L_y=2l$ and $L_z=D$, with $D$ (the wall separation) fixed and $l$ varied to change the system size. For the Binder-cumulant analysis, we considered two out of six slabs, each of size $l \times 2l \times D$, in the maximum- and minimum-density regions.

All simulations were performed using the GROMACS package (version 2023.2 or higher), with primary production runs carried out in the NVT ensemble using the Nose-Hoover thermostat\cite{van2005gromacs, nose2002molecular, hoover1985canonical}. The TIP4P/2005 water model was used throughout with a time step of 1\,fs and the LINCS algorithm was employed to maintain the geometry of water molecules\cite{hess1997lincs}. Periodic boundary conditions were applied in all three dimensions, except for simulations of confined water, where they were applied only in the $xy$ plane. To account for long-range Coulombic interactions, the 3d particle mesh Ewald (PME) method was utilized with a grid spacing of 1 Å for bulk and ionic systems. For confined systems, a quasi-two-dimensional Ewald summation method ("3dc" in GROMACS) was employed\cite{yeh1999ewald}. For the Lennard-Jones interactions, multiple studies have highlighted the significant impact of the cutoff value on the critical point \cite{smit1992phase, hu2015liquid, klauda2007long, qiu2019systematic, miguez2013influence, shen2007comparative, sega2017long, panagiotopoulos1994molecular, heyes2015lennard, blas2008vapor, macdowell2009surface}. Accordingly, we conducted simulations with cutoff values ranging from 9\AA~to 16\AA~for bulk water system. Long-range corrections to both energy and pressure were applied beyond the cutoff\cite{ allen2017computer}. For completeness, we also performed simulations employing full Lennard-Jones interactions, utilizing the LJ-PME method to account for long-range dispersion forces beyond the cutoff\cite{essmann1995smooth, xu2021validating, shirts2007accurate}. The typical system sizes ranged from $N=$ 1000 to 4000, and the box dimensions were chosen so that the average density was approximately \(0.31\,\mathrm{g/cm^3}\). Configurations were saved every few ps. However, for analysis, frames were selected at intervals determined by the relaxation time of the density autocorrelation function in each subbox, ensuring statistical independence between successive configurations. Simulations were typically run for several hundred nanoseconds, with some cases (particularly for larger systems near the critical point) extending up to a few microseconds to ensure proper equilibration and adequate sampling. Fig. 1b shows the cumulant crossing for the LJ-PME case.

To assess the robustness of the subbox-based NVT methodology described above, we performed additional simulations under the NPT ensemble for the LJ-PME case using the Parrinello--Rahman barostat with different system sizes\cite{parrinello1981polymorphic}. In brief, for each system size, $T_c(L)$ was estimated by fitting the fluctuations of a mixed order parameter $t = \rho + sE$ to the universal probability distribution of the 3D Ising model. Distributions $P(t)$ were generated via histogram reweighting, then shifted and scaled to obtain a normalized variable $M = (t - \bar{t})/\sigma_t$. The resulting distribution $P(M)$ was compared to the universal Ising form using a least-squares error metric, and the optimal $T$, $P$, and $s$ were determined using the \textsc{Minuit} minimizer, yielding estimates of $T_c$, $P_c$, and $s$\cite{debenedetti2020second, tsypin2000probability}. Standard finite-size scaling was then applied using the relation
\begin{equation}
    T_c(L) = T_c + aL^{-1/\nu}
\label{eq:4}
\end{equation}
where $a$ is a proportionality constant and $\nu$ is the critical exponent for the correlation length. We fixed $\nu = 0.629912$, corresponding to the 3D Ising universality class\cite{ferrenberg2018pushing}, and determined $T_c$ and $a$ by minimizing the least-squares error.

\section{Results}

\subsubsection{Critical Point of Bulk Water}

\begin{figure}[t]
  \centering
  \begin{minipage}[c]{0.45\linewidth}
    \centering
    \includegraphics[width=\linewidth]{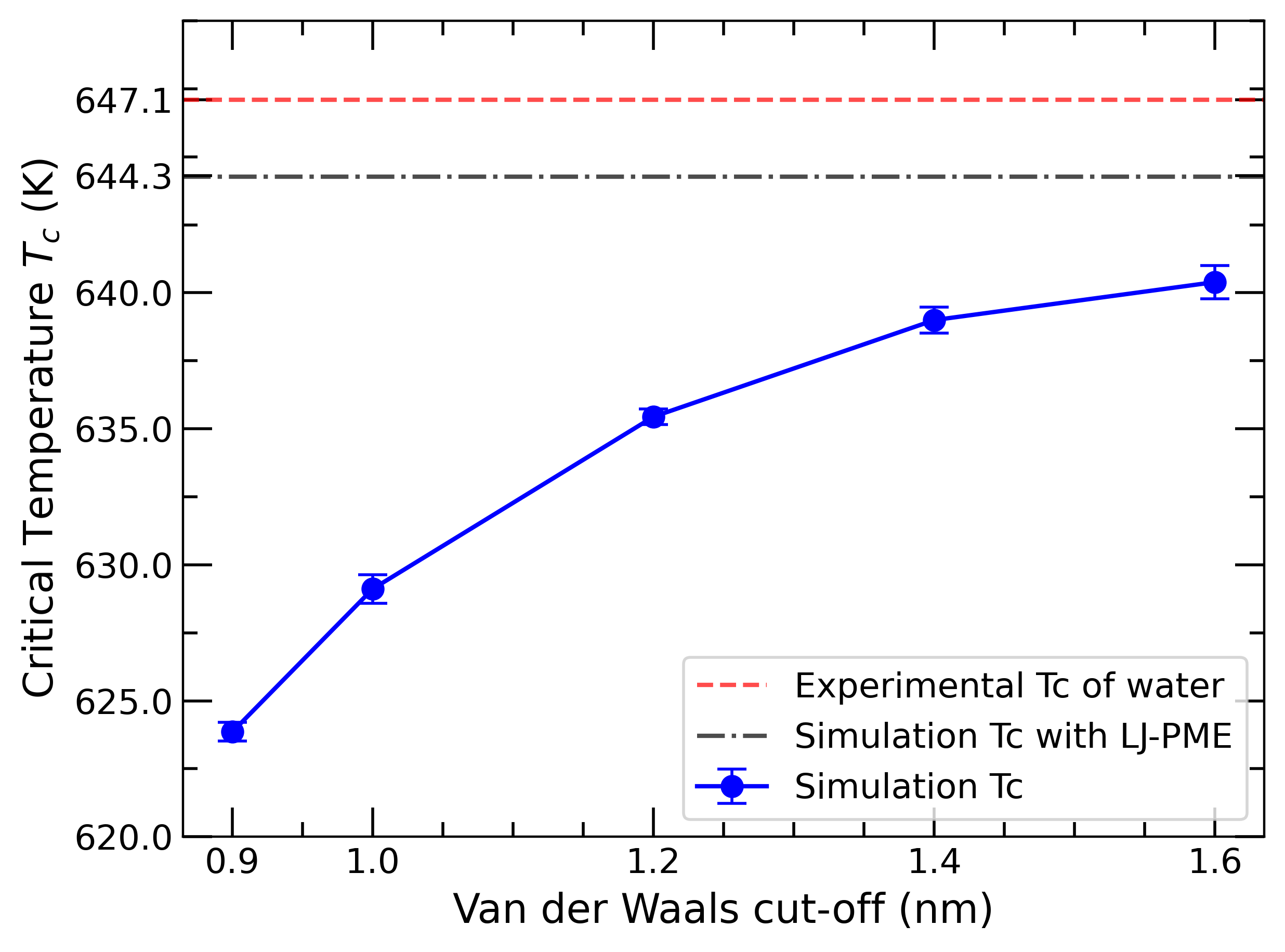}
    \par\smallskip
    (a)
  \end{minipage}
  \hfill
  \begin{minipage}[c]{0.45\linewidth}
    \centering
    \includegraphics[width=\linewidth]{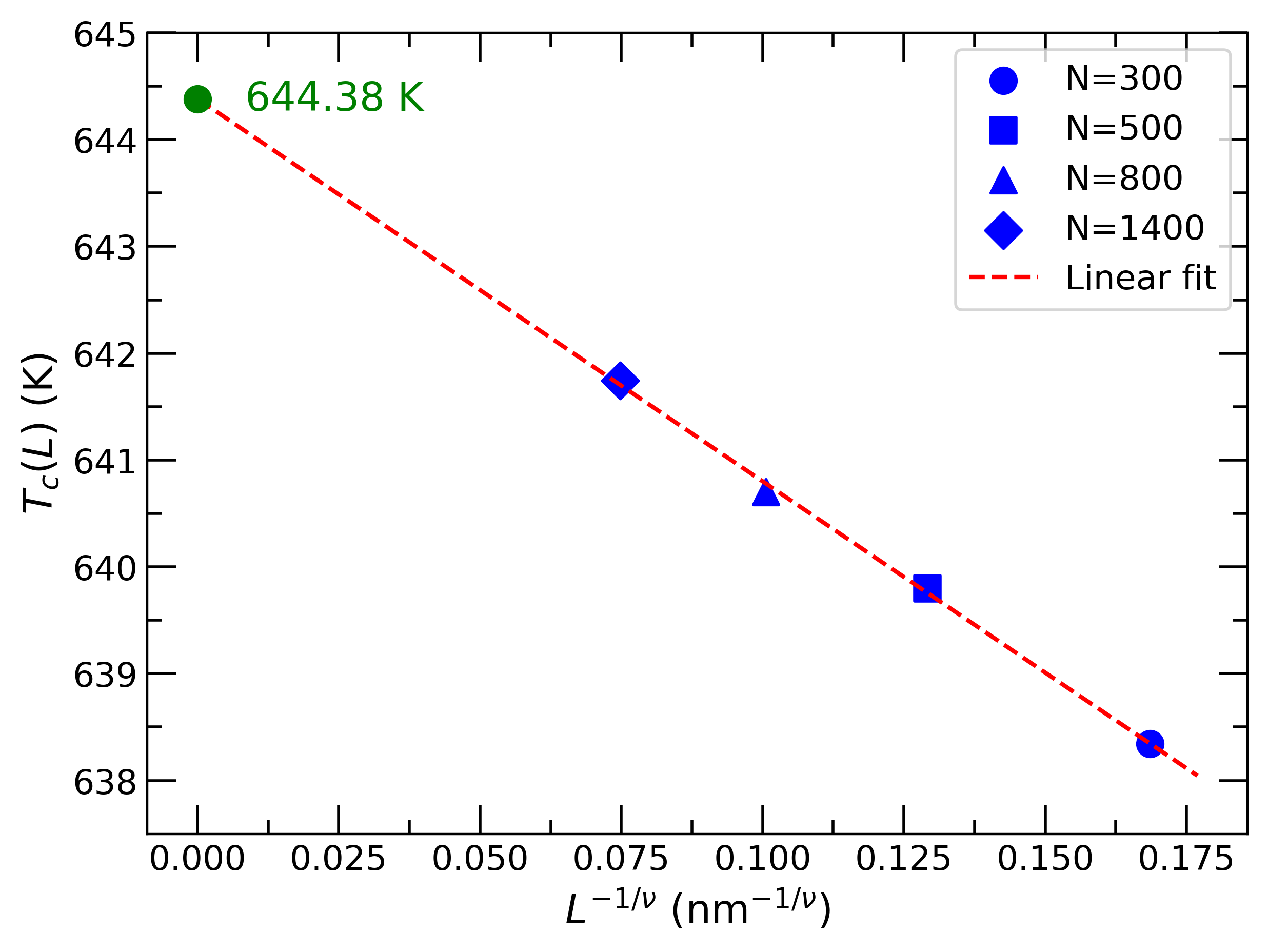}  
    \par\smallskip
    (b)
  \end{minipage}
  \caption{
  (a) Dependence of the estimated critical temperature of bulk water on the Lennard--Jones (LJ) cutoff length. Even with long-range corrections applied, short LJ cutoffs significantly underestimate the critical temperature (by up to 20\,K), while increasing the cutoff shifts $T_c$ upward toward the value obtained with the LJ-PME method. 
  (b) Finite-size dependence of the critical temperature obtained from NPT simulations for $N=300$, 500, 800, and 1400. The estimated $T_c$ increases with system size, and extrapolation using the first-order finite-size scaling relation yields a thermodynamic-limit value of $T_c \approx 644.3$\,K.}
  \label{fig:2}
\end{figure}
 
Figure~\ref{fig:2}a illustrates the impact of the Lennard-Jones cutoff on the critical point. To isolate the effect of the Lennard-Jones cutoff, the real-space cutoff for the Coulombic interactions was fixed at 16\AA~in all simulations, with long-range electrostatics consistently treated using the PME method as described in the Methods section. Significant deviations in the critical temperature are observed when shorter Lennard--Jones cutoffs are used, leading to an underestimation of $T_c$ by up to 20~K compared to the LJ-PME approach. As the cutoff increases, the critical point systematically shifts toward higher temperatures, approaching the LJ-PME value, but even with a cutoff of 1.6~nm, $T_c$ is still underestimated by about 4~K. Our results are in close agreement with those reported in Ref.~\onlinecite{hu2015liquid}, although our estimates are consistently higher by about 3--5~K. This discrepancy may arise from methodological differences; notably, the authors of Ref.~\onlinecite{hu2015liquid} primarily estimated the critical temperature using a scaling law fit to the liquid--vapor density difference, and also employed analytical fits to temperature-dependent surface tension as a consistency check. Our method yields a critical temperature of 644.3 $\pm$ 0.5~K for the LJ-PME case (Fig.~1b), in the immediate vicinity of the experimental value of 647.1~K. Although long-range corrections were applied for truncated Lennard--Jones cutoffs, deviations in the critical temperature persisted relative to the LJ-PME reference, emphasizing the limitations of truncation-based methods.  

To further validate the robustness of the subbox-based NVT approach, we additionally performed NPT simulations for several system sizes to estimate the critical point. Figure~\ref{fig:2}b shows the critical temperatures obtained for $N = 300$, 500, 800, and 1400. While Ref.~\onlinecite{debenedetti2020second} estimated the critical parameters by averaging results from a few system sizes, we have explicitly accounted for finite-size dependence and extrapolated to the thermodynamic limit using the first-order finite-size scaling ansatz (Eq.~\ref{eq:4}). Pronounced finite-size effects are observed: the estimated $T_c$ systematically increases with system size, with the smallest system yielding a value approximately 6~K lower than the extrapolated thermodynamic-limit value of $T_c \approx 644.4$~K, as calculated using Eq.~\ref{eq:4}. First-order finite-size scaling provides a suitable description for the system sizes considered here and generally captures the dominant finite-size trends in NPT-based estimates of critical parameters, although higher-order corrections may affect very small simulation boxes~\cite{ferrenberg2018pushing}. The estimated thermodynamic-limit value is in excellent agreement with the critical temperature determined independently from the subbox-based NVT approach confirming that both approaches reliably capture the critical behavior and motivating the use of the NVT method for the analysis in the following section.

\subsubsection{Effect of Ions and Confinement on the Critical Point of Water}

To investigate the influence of dissolved ions on the critical behavior of water, we employed the MADRID-2019 force field, which was developed for compatibility with the TIP4P/2005 water model. In these simulations, the Lennard-Jones cutoff was fixed at 1~nm and long range corrections of the potential energy and pressure were included, as in the original force field implementation\cite{zeron2019force}. The subbox-based sampling methodology described earlier was applied to determine the critical point in NaCl solutions at varying concentrations, where only water oxygen atoms were sampled; the ions’ contribution to the total density is negligible at these concentrations.  

\begin{figure}[t]
  \centering
  \begin{minipage}[c]{0.45\linewidth}
    \centering
    \includegraphics[width=\linewidth]{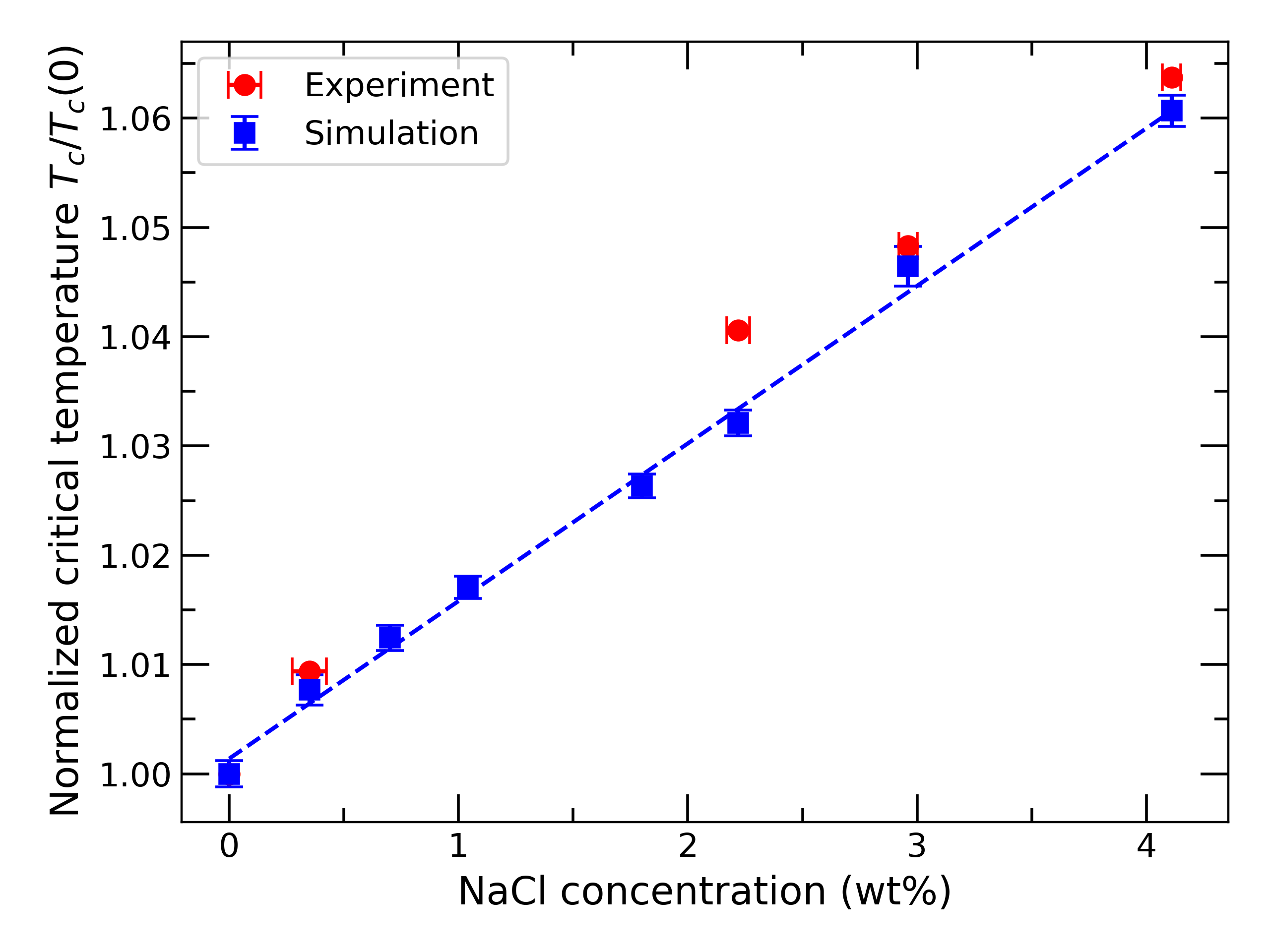}
    \par\smallskip
    (a)
  \end{minipage}
  \hfill
  \begin{minipage}[c]{0.45\linewidth}
    \centering
    \includegraphics[width=\linewidth]{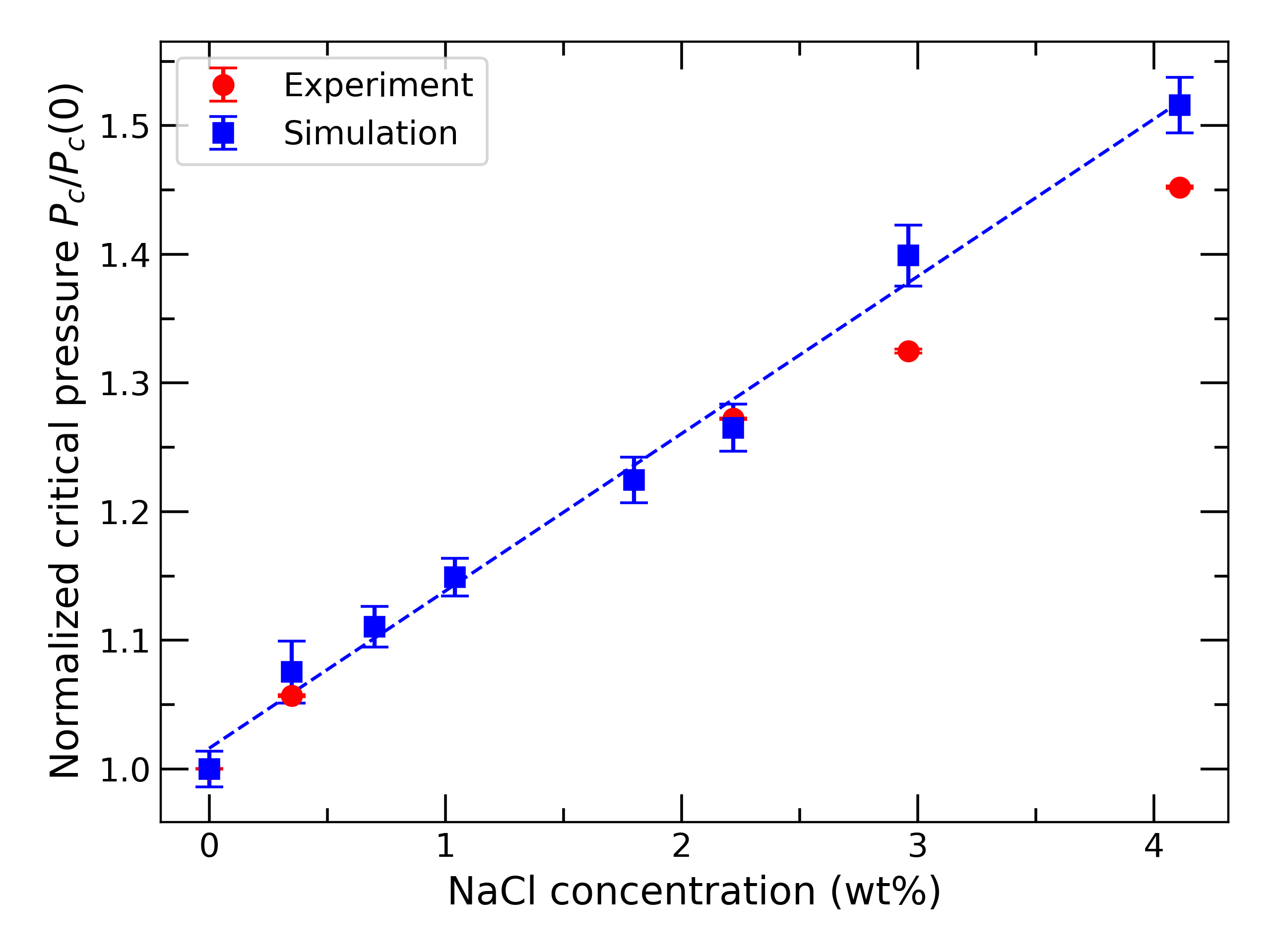}
    \par\smallskip
    (b)
  \end{minipage}
  \caption{(a) Normalized critical temperature $T_c(x)/T_c(0)$ and (b) normalized critical pressure $P_c(x)/P_c(0)$ as functions of NaCl concentration $x$ (wt\%). Simulation results (blue) are compared with experimental measurements (red) from Ref.~\onlinecite{bischoff1988liquid}. Both critical parameters increase linearly with salt concentration. The dashed blue line is the linear fit to the simulation data.}
  \label{fig:3} \
\end{figure}

Figure \ref{fig:3} shows the normalized shifts in the critical temperature (a) and critical pressure (b) with increasing NaCl concentration. Both the $T_c$ and $P_c$ exhibit a clear monotonic rise, consistent with the established experimental studies of aqueous electrolyte solutions~\cite{bischoff1988liquid, marshall1990critical, povodyrev1999critical, bischoff1989liquid}. This upward shift originates from well-known microscopic effects of dissolved ions on near-critical water: strong ion--water interactions generate electrostriction and structured hydration shells, locally increasing the density of the liquid\cite{marshall1993volume, cochran1992solvation}. Thermodynamic analyses show that the solution reaches its critical point when the density of the remaining “free” (unsolvated) water approaches the critical density of pure water~\cite{marshall1990critical}, reflecting the fact that hydration shells effectively remove water from participating in large-scale phase separation. As the system approaches criticality, these dense hydration regions interact with and amplify long-wavelength density fluctuations, producing denser and larger fluctuations than in pure water. Such ion-induced enhancement of density fluctuations has been observed experimentally and discussed in studies of ionic fluids near their critical region~\cite{onuki2004solvation, testemale2005small}. Although the Madrid-2019 force field was not parametrized explicitly to reproduce phase diagrams of salt solutions, it nevertheless provides consistent agreement with experimental data, along with rather accurate quantitative relative shifts in the critical temperature across the studied concentration range. Note, however, that while the critical temperature for TIP4P/2005 is close to the experimental value, the absolute value of the critical pressure is typically underestimated.

\begin{figure}[t]
  \centering
  \includegraphics[width=0.6\linewidth]{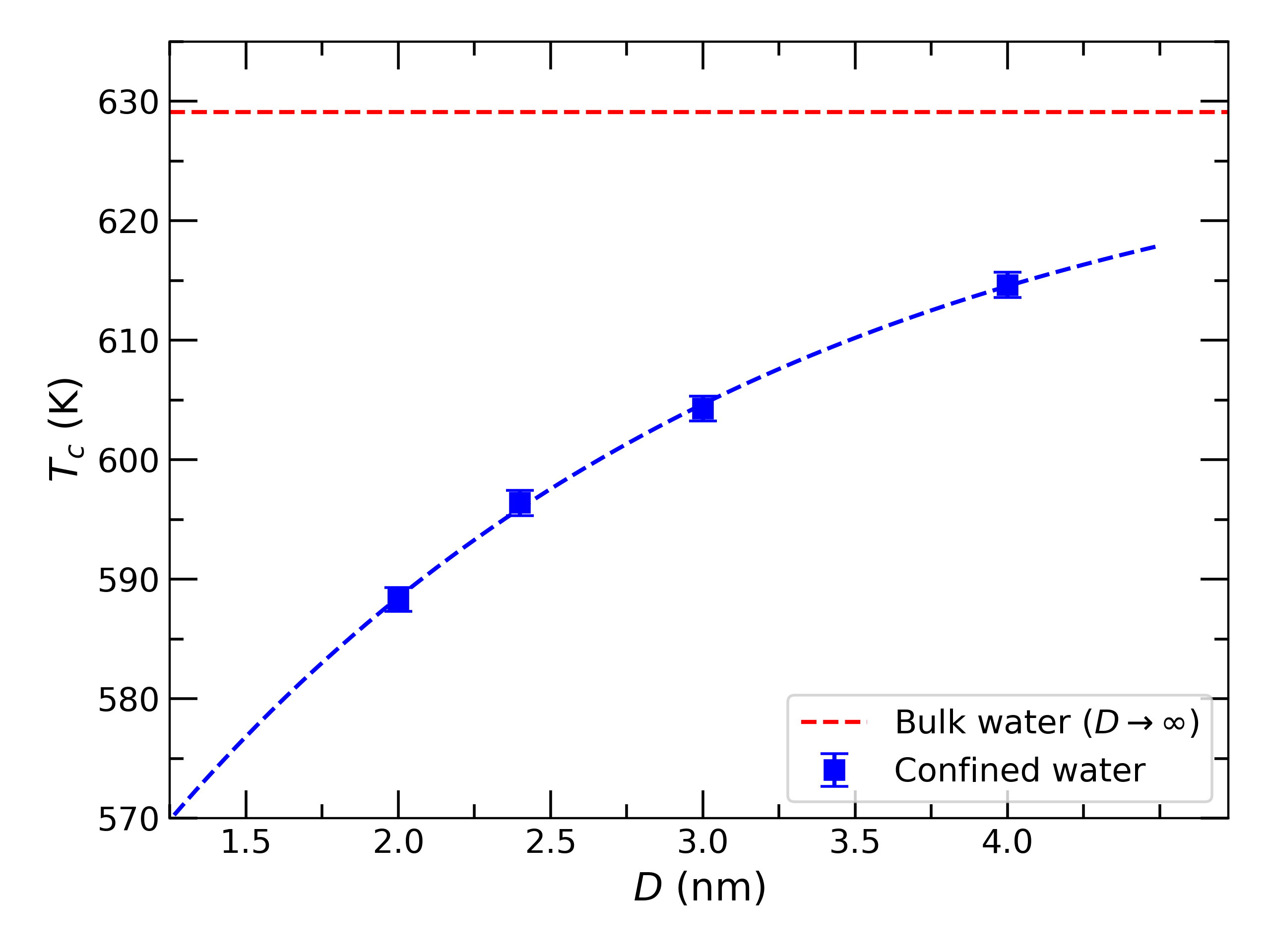}  % adjust width as needed
  \caption{Critical temperature $T_c$ of water confined between hydrophobic walls, plotted against the slit width $D$. Simulation values (blue squares) were determined using subbox-based density fluctuations. The red dashed line marks the bulk $T_c$ obtained under identical simulation conditions (LJ cutoff of 1~nm). The dashed blue line is a guide to the eye.}
  \label{fig:4}
\end{figure}

For confined water systems, we analyzed the shifts in the critical temperature using subbox-based density fluctuations, focusing on regions minimally affected by the walls. Using the Barker--Henderson formula~\cite{barker1967perturbation, zausch2009statics}, the effective hard-sphere diameter of the wall was estimated to be $d \approx 0.095$~nm. Here, the effective confinement width is defined as $D_{\rm eff} = D - 2d$, where $d$ is the effective hard-sphere radius of the wall. The van der Waals cutoff was set to 1.0~nm. As shown in Figure~\ref{fig:4}, confinement leads to a systematic depression of $T_c$ relative to bulk water, with the magnitude of the shift increasing as the effective lateral system size is reduced. The red dashed line is the bulk water critical temperature ($\approx 629$~K as seen in Figure~\ref{fig:2}a) for the Lennard-Jones cutoff of 1~nm. The trend highlights the influence of hydrophobic confinement on the depression of the liquid--vapor critical temperature, consistent with prior simulation studies~\cite{brovchenko2001phase, brovchenko2004water, brovchenko2004water2}. 

\section*{Conclusion}

We performed a rigorous estimation of the liquid--gas critical point of the TIP4P/2005 water model and investigated how it is affected by salt (NaCl) and confinement.

To ensure methodological accuracy, we first determined $T_c$ for bulk water using the particle--mesh Ewald (PME) treatment of Lennard--Jones interactions. Two complementary approaches were employed: (i) a recently developed subbox-based Binder cumulant method in the NVT ensemble, originally applied to active Brownian particle systems, and (ii) histogram reweighting of joint density--energy distributions from NPT simulations, fitted to the three-dimensional Ising universal scaling curve. Both yielded consistent critical temperatures, confirming the reliability of our approach. We also assessed the sensitivity of $T_c$ to the Lennard--Jones cutoff distance in NVT simulations, finding a systematic increase from approximately 623~K (0.9~nm cutoff) to about 644~K (LJ-PME).

Using the validated NVT approach, we studied aqueous NaCl solutions with the Madrid-2019 force field. The results show a systematic increase in the critical temperature and pressure with salt concentration, consistent with experimental trends. Quantitative deviations are mainly attributed to the 1.0~nm Lennard--Jones cutoff prescribed in the Madrid-2019 model, which yields a bulk water $T_c \approx 629$~K. When normalized by this bulk value, the variation of $T_c$ with concentration agrees well with experiments.

Finally, simulations of water confined between hydrophobic walls revealed a depression of $T_c$, consistent with previous studies. The methodology developed and tested in this study can be readily applied to investigate the liquid--liquid critical point of water and to assess the validity of machine-learning--based force fields trained on quantum-mechanical data. Our MD-based approach can also be adapted to other complex liquids for which grand canonical Monte Carlo simulations are difficult.

\begin{acknowledgments}
Funding from the DFG (Deutsche Forschungsgemeinschaft) is acknowledged: P.V. is a member of the GRK 2516 (project No. 405552959) and M.S. is the recipient of a doctoral position within the GRK 2516 program. In addition, we would like to thank Luis Gonzalez MacDowell and Łukasz Baran for kindly sharing reference wall configurations used in their previous studies. We are also grateful to Katrin Amann-Winkel, Isabell Zick, Yizhi Liu and Tobias Eklund. The authors acknowledge the use of ChatGPT (OpenAI) for assistance with language and text refinement. Computational resources were provided by the supercomputers MOGON II and MOGON III at Johannes Gutenberg University Mainz, which are part of the NHR South-West infrastructure.
\end{acknowledgments}

\section*{Author Declarations}

\textbf{Conflict of Interest:} The authors have no conflicts to disclose.

\textbf{Author Contributions:} 
Mayank Sharma: Data curation (lead); Formal analysis (lead); Investigation (lead); Methodology (lead); Software (lead); Visualization (lead); Writing – original draft (lead); Writing – review \& editing (supporting).  
Peter Virnau: Conceptualization (lead); Funding acquisition (lead); Investigation (supporting); Methodology (supporting); Project administration (lead); Supervision (lead); Visualization (supporting); Writing – original draft (supporting); Writing – review \& editing (lead).

\section*{Data Availability Statement}

The data of the figures will be uploaded to Zenodo. Further data is available from the corresponding author upon reasonable request.

\bibliography{aipsamp}% Produces the bibliography via BibTeX.

\end{document}